\documentclass[print]{aa}

\usepackage{times}
\usepackage[round]{natbib}
\usepackage{pdflscape}
\usepackage{afterpage}
\usepackage{capt-of}
\usepackage{color}
\usepackage{hyperref}
\usepackage{graphicx}
\usepackage{longtable}
\usepackage{tabu}
\usepackage{color,soul}
\usepackage{multirow}
\usepackage{amsmath}
\usepackage{verbatim}
\usepackage{pdflscape}
\usepackage{caption}
\usepackage{subcaption}
\usepackage{amsmath}
\usepackage[normalem]{ulem}
\usepackage[autostyle]{csquotes}
\usepackage{siunitx}
\usepackage{float} 
\usepackage{url}

\hypersetup{draft}
\begin{document}

\title{Dynamical evolution of basaltic asteroids\\ outside the Vesta family in the inner main belt}

\titlerunning{Dynamical evolution of basaltic asteroids}
\authorrunning{Troianskyi et al.}

\author{Volodymyr Troianskyi \inst{\ref{inst1},\ref{inst2},\ref{inst3}} \and Pawe\l{} Kankiewicz \inst{\ref{inst4}} \and Dagmara Oszkiewicz \inst{\ref{inst1}}}

 \institute{Astronomical Observatory Institute, Faculty of Physics, Adam Mickiewicz University, S{\l}oneczna 36, 60-286 Pozna{\'n}, Poland \label{inst1} \and
 Department of Physics and Astronomy FMPIT of Odesa I. I. Mechnykov National University, Pastera Street 42, 65082 Odesa, Ukraine \label{inst2} \and
 Astronomical Observatory of Odesa I. I. Mechnykov National University, Marazlievskaya 1v, 65014 Odesa, Ukraine \label{inst3} \and
 Institute of Physics, Jan Kochanowski University, Uniwersytecka 7, 25-406 Kielce, Poland \label{inst4}
}
         
\date{\today} 

\offprints{V. Troianskyi, e-mail: voltro at amu.edu.pl}

\date{Received xx xx xxxx / Accepted xx xx xxxx} 
   
 \abstract
 {{
 Basaltic V-type asteroids are leftovers from the formation and evolution of differentiated planetesimals. They are thought to originate from mantles and crusts of multiple different parent bodies. Identifying the links between individual V-type asteroids and multiple planetesimals is challenging, especially in the inner part of the main asteroid belt, where the majority of V-type asteroids are expected to have originated from a single planetesimal, namely, (4) Vesta.
 }}
 {{
 In this work, we aim to trace the origin of a number of individual V-type asteroids from the inner part of the main asteroid belt. The main goal is to identify asteroids that may not be traced back to (4) Vesta and may therefore originate from other differentiated planetesimals.
 }}
 {{
 We performed a 2 Gy backward numerical integration of the orbits of the selected V-type asteroids. For each asteroid, we used 1001 clones to map the effect of orbital uncertainties. In the integration, we use information on physical properties of the considered V-type asteroids such as pole orientation, rotational period, and thermal parameters.
 }}
 {{
The majority of the studied objects can be traced back to the Vesta family within 2 Gy of integration. The number of objects of the low-inclination V-types did not reach the boundary of the Vesta family during the integration time. Two asteroids, namely, (3307) Athabasca and (17028) 1999 FJ$_{5}$, do not show a dynamic link to (4) Vesta. Increasing the integration time for these objects leads to further separation from (4) Vesta.
 }}
 {{
 The majority of V-types in the inner main belt outside the Vesta family are clearly Vesta fugitives. Two objects, namely, (3307) Athabasca and (17028) 1999 FJ$_{5}$, show no clear dynamical link to (4) Vesta. Together with (809) Lundia (from our previous work), these objects could represent the parent bodies of anomalous HED meteorites such as the Banbura Rockhole. Furthermore, some objects of the low-inclination population cannot be traced back to (4) Vesta within the 2 Gy integration.
 }}


\maketitle

\section{Introduction}
\label{intro}

Generally, V-type asteroids are known to trace the history of differentiated planetesimals in the Solar System \citep{nesvorny2008fugitives}. These bodies were the precursors of terrestrial planets and thus hold clues to planetary formation and the more general evolution of the Solar System. Specifically, the number of differentiated planetesimals and their distribution map the evolution of our planetary system \citep{burbine2002meteoritic, Scott2015Asteroid}. The planetesimal formation and evolution theory by \cite{bottke2006iron, bottke2014origin} suggest that these bodies formed close to the Sun in the terrestrial planet region. Other authors have suggested wider formation location ranges (1.3 au - 7.5 au) \citep{lichtenberg2021bifurcation}. These bodies were then collisionally disrupted and scattered into the main asteroid belt. Some of the fragments were then recovered on Earth as meteorites. In particular, iron meteorites, which are believed to originate from the iron cores of differentiated planetesimals, suggest that there were 100-150 such bodies in the early Solar System \citep{burbine2002meteoritic}. Fragments of those disrupted bodies should still be plentiful in the Main Asteroid Belt, especially in the inner section (a $<$2.5 au), which is dynamically easier to evolve from the terrestrial planet region and from which most meteorites come \citep{bottke2006iron}. 

However, up-to-date spectral observations do not show a large number of distinct V-types (parts of mantles of the differentiated bodies) across the Solar System. Most V-types reside in the inner main belt in the vicinity of the fossil planetesimal (4) Vesta and are therefore considered related \citep{binzel1993chips, bus2002phase, DeMeo2009extension, moskovitz2008spectroscopically, moskovitz2010spectroscopic, Popescu2018Taxonomic, oszkiewicz2019physical, oszkiewicz2020spin}.  Most are parts of the Vesta family or considered Vesta fugitives, that is, objects that escaped the borders of the family through the combination of the Yarkovsky effect and dynamical resonances \citep{Nesvorny2015HCM, nesvorny2008fugitives}. Objects genetically related to (4) Vesta are commonly named Vestoids and those that cannot be linked to (4) Vesta are known as non-Vestoids. There is strong evidence linking (4) Vesta, Vestoids, and Howardite-Eucrite-Diogenite meteorites (HEDs), including those delivered by the NASA Dawn mission. This link between HEDs, Vestoids, and (4) Vesta was first identified in 70 ties through spectral observations \citep{mccord1970asteroid}. Subsequent observations of other V-types that extended across the 3:1 and $\nu_6$ Jovian resonances have provided a plausible Earth-delivery scenario \citep{binzel1993chips, burbine2001vesta}. More detailed comparative petrologic and geochemical measurements of the Vesta's surface and HEDs further strengthened that link \citep{mcsween2013dawn}. The two largest craters on the surface of (4) Vesta are thought to be the main source of most Vestoids and HEDs \citep{thomas1997impact, Marchi2012, schenk2012geologically}. The age of those craters (Rheasilvia and Veneneia) are thought to be ~1Gy and ~2Gy, respectively, and correspond to the age of the Vesta family \citep{schenk2022young, spoto2015}. Taking into account the above considerations described in the references cited above, we chose an integration time of 2 Gy for our dynamic simulation described below, as best comparable to the age of the Vesta family.

Some unique V-types were identified beyond the 3:1 and 5:2 mean motion resonances with Jupiter. The first of those was (1459) Magnya, which was recognized as basaltic in the early 2000s \cite{lazzaro2000discovery}. Based on dynamical investigations, \cite{michtchenko2002origin} suggested that this $\sim$30 km asteroid located beyond 2.8 au from the Sun is most probably a part of other (than (4) Vesta) planetesimal that existed in the outer parts of the main belt. \cite{hardersen2004mineralogy} further substantiated this claim by finding the discordant pyroxene chemistry of (1459) Magnya compared to that of (4) Vesta. Currently, a dozen other V-type asteroids have been identified in the middle and outer parts of the main belt \citep{duffard2009two, hammergren200621238, ieva2016spectral, ieva2018basaltic}. \cite{roig2008v} has shown that large asteroids ($>$ 5 km) in the middle of the main belt region have a low probability $\sim$1\% to have evolved from (4) Vesta through a combination of the Yarkovsky effect and dynamical resonances. \cite{ieva2016spectral} and \cite{leith2017compositional} analyzed their spectroscopic and mineralogical properties and found surface compositions that are not compatible with that of (4) Vesta. \cite{ieva2018basaltic} suggested that the Eos asteroid family in the outer main belt could be the source of some of the V-types in this region. The dynamical evolution of V-type candidates also suggests that the parent bodies of the Eunomia and Merxia / Agnia families could also be a potential source of V-types in the middle and outer main belt \citep{carruba2014dynamical}. \cite{huaman2014dynamical} identified three possible source regions of the V-types in the outer main belt, associated with the parent bodies of (1459) Magnya, (349) Debowska, and (221) Eos. Objects in the middle and outer main belt could also be delivered to its current locations from the inner main belt through the so-called "jumping Jupiter" mechanism \citep{brasil2017scattering, Migliorini2021Characterization}.

Theoretically, the number of V-type fragments originating from distinct planetesimals should be even greater in the inner main belt \citep{bottke2006iron}. However, most of the V-type asteroids in the inner belt of the main belt are parts of the Vesta family or are considered fugitives \citep{nesvorny2008fugitives}, which are objects that evolved away from the family and are now beyond recognition as family members with traditional clustering methods \citep{Nesvorny2015Asteroid}. Early studies showed that some of the V-types in the inner main belt present deeper 1.0 $\mu$m absorption bands than (4) Vesta \citep{Florczak2002Discovering}. However, these findings could not be confirmed by \cite{ieva2016spectral, ieva2018basaltic}. Recently, \cite{OszkiewiczGaia2023} showed that most V-type asteroids in the inner main belt have spectral properties that overlap with that of the Vesta family. However, few V-type asteroids were found with an unusually deep 0.9~$\mu$m band depth, which is more consistent with those of (1459) Magnya \citep{lazzaro2000discovery}; thus, deeper mineralogical analysis is required. 

There is limited evidence that in addition to a large population of Vestoids, there might be some non-Vestoids present in the inner main belt. \cite{bland2009anomalous} and \cite{spurny2012bunburra} observed a fall of anomalous HED meteorite (V-type material not related to (4) Vesta) and estimated  there was a very high probability (of 98\%)  that
it had originated  from the innermost main belt. \cite{oszkiewicz2015differentiation} showed that due to its observationally constrained prograde rotation (809), Lundia is unlikely to be a former Vesta family member. Earlier dynamical work on (809) Lundia suggested a link to (4) Vesta, but did not consider the prograde rotation of the object \citep{carruba2005v}.

In this work, we investigate a number of V-type asteroids in the inner main belt with known spin axis coordinates. We performed a numerical integration on the 2 Gy scale (estimated age of the Vesta family \citep{spoto2015, schenk2012geologically}) to investigate their possible origin and verify the hypothetical presence of non-Vestoids in the inner main belt. In Sect. \ref{targets}, we report the objects studied. In Sect. \ref{dynam}, we describe the dynamical model used in this study. The results are presented in Sect. \ref{res}, along with a discussion of our results in Sect. \ref{dis} and conclusions in Sect. \ref{summ}.

\section{Target selection}
\label{targets}
We selected V-type asteroids (identified spectrally or by multifilter photometry) outside the dynamical Vesta family and for which spin and shape are determined. The family membership was extracted from \cite{Nesvorny2015HCM}. The selected objects are listed in Table \ref{tab:targets}. Detailed knowledge of the sidereal period and the sense of rotation (along with the size and other parameters) determines the direction and scale of the Yarkovsky drift and is incorporated into into the dynamical integration. 

\begin{figure}
    \centering
    \includegraphics[width=0.5 \textwidth]{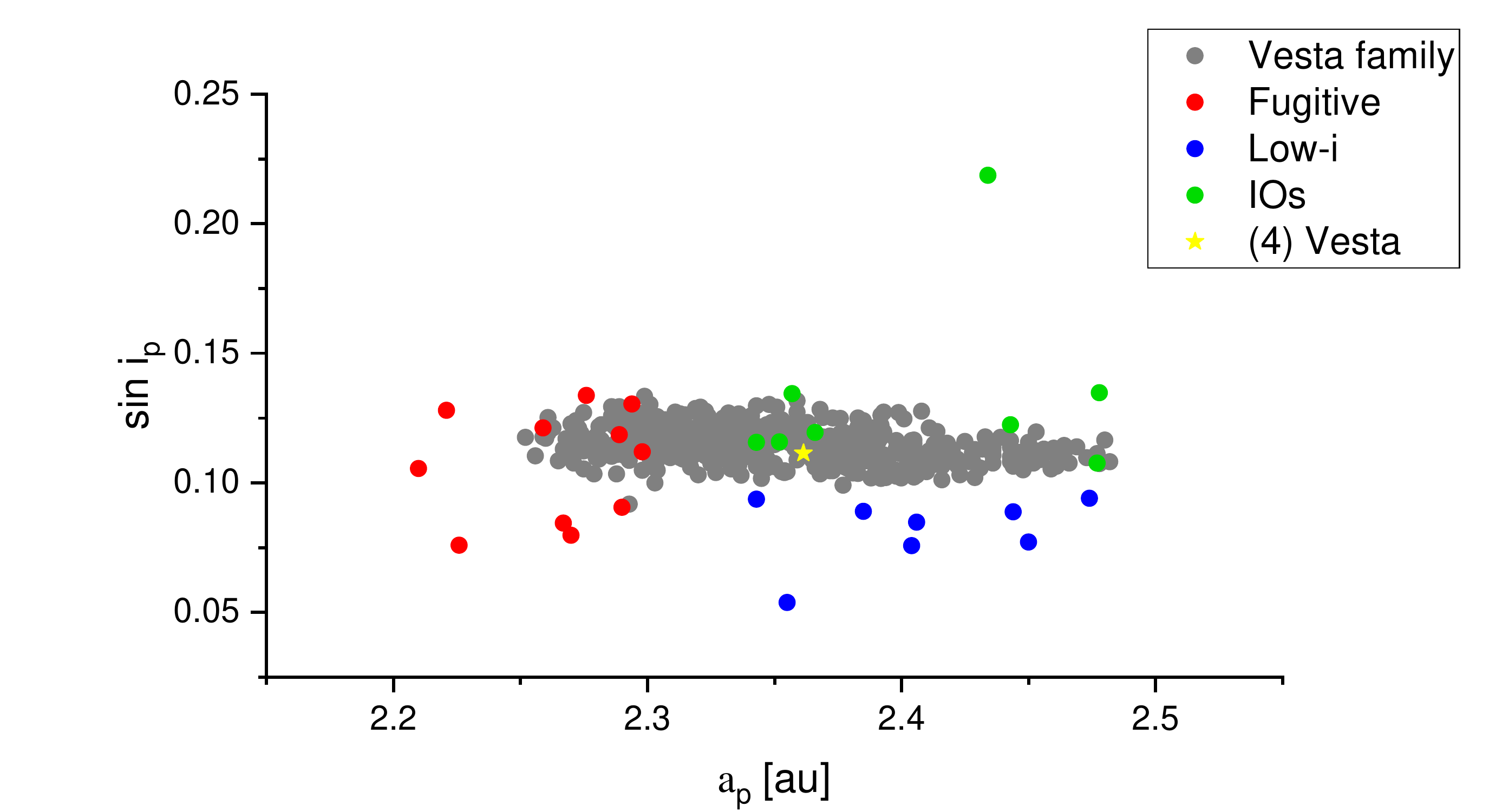}
    \caption{Orbital distribution of asteroids considered in this work.}
    \label{i_a}
\end{figure}

We further divided the objects into categories based on orbital elements (see Table \ref{tab:targets} and Fig. \ref{i_a}): Fugitives - objects outside the dynamical Vesta family having $2.1\,\text{au} < a \le 2.3\,\text{au}, 5~\degr < i < 8~\degr$, and $0.035 < e < 0.162$; Low-$i$ - objects outside the dynamical Vesta family having $2.3\,\text{au} < a \le 2.5\,\text{au}$ and $i < 6~\degr$; Inner other - remaining objects in the inner main belt outside the dynamical Vesta family.

 The division into fugitives, low-$i$, and inner other is consistent with previous spectral and dynamical studies of V-types \citep{ieva2016spectral, ieva2018basaltic, nesvorny2008fugitives, OszkiewiczGaia2023}. Fugitives are objects with a smaller semimajor axis than Vesta family members and comparable orbital inclination and eccentricity. In principle, those objects should be easily explained by migration from the Vesta family. Low-inclination objects (with orbital inclinations smaller than typical members of the Vesta family) could not be previously  fully explained (in a dynamical sense) based on the assumption of an impact on (4) Vesta 2 Gy ago. Furthermore, lithological differences between Vesta family members and low-$i$ asteroids were reported by \cite{mansour2020distribution}. \cite{nesvorny2008fugitives} suggest that these objects originated in an earlier impact on (4) Vesta, plausibly 3.9 Gy ago, or originated from a different differentiated parent body. The last population are the remaining V-types in the inner main belt.
 
\begin{table}[h]
\caption{\label{tab:targets}Asteroids covered in this study.}
    \centering
    \scriptsize\addtolength{\tabcolsep}{-2pt}
    \begin{tabular}{l|c|c|c|c|c|c}
    \hline
    \hline
          Asteroid No. & H & $a$ & $e$ & sin ($i$) & Sense of & Notes \\
          and name & & [au] & & & rotation & \\
        \hline
          (956) Elisa & 12.13 & 2.298 & 0.158 & 0.1119 & ret.$^{[1,2]}$ & Fugitive \\
      \hline
          (1914) Hartbeespoortdam & 12.09 & 2.406 & 0.139 & 0.0848 & pro.$^{[2]}$ & Low-i \\
          \hline
          (1946) Walraven & 12.13 & 2.294 & 0.190 & 0.1304 & ret.$^{[3,4]}$ & Fugitive \\
          \hline
          (2432) Soomana & 12.76 & 2.352 & 0.129 & 0.1158 & pro.$^{[1,2]}$ & IOs \\
          \hline
          (2566) Kirghizia & 12.48 & 2.450 & 0.104 & 0.0771 & ret.$^{[1,2]}$ & Low-i \\
          \hline
          (2579) Spartacus & 13.59 & 2.210 & 0.082 & 0.1054 & ret.$^{[5]}$ & Fugitive \\
          \hline
          (2653) Principia & 12.22 & 2.444 & 0.114 & 0.0888 & pro.$^{[1,2]}$ & Low-i \\
          \hline
          (2704) Julian Loewe & 12.81 & 2.385 & 0.117 & 0.0889 & ret.$^{[2,4]}$ & Low-i \\
          \hline
          (2763) Jeans & 12.47 & 2.404 & 0.179 & 0.0756 & ret.$^{[2]}$ & Low-i \\
          \hline
          (2851) Harbin & 12.32 & 2.478 & 0.123 & 0.1348 & pro.$^{[2,7]}$ & IOs \\ 
          \hline
          (2912) Lapalma & 12.76 & 2.289 & 0.118 & 0.1186 & ret.$^{[2]}$ & Fugitive \\
          \hline
          (3307) Athabasca & 14.12 & 2.259 & 0.100 & 0.1212 & pro.$^{[1]}$ & Fugitive \\
          \hline
          (3536) Schleicher & 14.02 & 2.343 & 0.077 & 0.1156 & ret.$^{[1,6]}$ & IOs \\
          \hline
          (3849) Incidentia & 13.09 & 2.474 & 0.065 & 0.0940 & pro.$^{[2]}$ & Low-i \\
          \hline
          (4796) Lewis & 13.65 & 2.355 & 0.141 & 0.0538 & ret.$^{[2,4]}$ & Low-i \\
          \hline
          (5150) Fellini & 13.32 & 2.477 & 0.138 & 0.1076 & pro.$^{[1]}$ & IOs \\
          \hline
          (5524) Lecacheux & 13.01 & 2.366 & 0.059 & 0.1194 & ret.$^{[1,2]}$ & IOs \\
          \hline
          (5525) 1991 TS$_{4}$ & 13.27 & 2.221 & 0.081 & 0.1279 & ret.$^{[2]}$ & Fugitive \\
          \hline
          (5754) 1992 FR$_{2}$ & 12.96 & 2.267 & 0.091 & 0.0843 & ret.$^{[2,4,6,7]}$ & Fugitive \\
          \hline
          (5952) Davemonet & 13.66 & 2.270 & 0.141 & 0.0796 & ret.$^{[1,6]}$ & Fugitive \\
          \hline
          (6406) Mikejura & 13.59 & 2.276 & 0.124 & 0.1336 & ret.$^{[2,3,5,8]}$ & Fugitive \\
          \hline
          (7558) Yurlov & 13.58 & 2.290 & 0.110 & 0.0904 & ret.$^{[2]}$ & Fugitive \\
          \hline
          (7899) Joya & 13.74 & 2.343 & 0.114 & 0.0937 & ret.$^{[6]}$ & Low-i \\
          \hline
          (17028) 1999 FJ$_{5}$ & 15.20 & 2.226 & 0.152 & 0.0758 & pro.$^{[6]}$ & Fugitive \\
          \hline
          (18641) 1998 EG$_{10}$ & 14.01 & 2.357 & 0.097 & 0.1344 & pro.$^{[4]}$ & IOs \\
          \hline
          (25327) 1999 JB$_{63}$ & 14.03 & 2.434 & 0.186 & 0.2187 & ret.$^{[1]}$ & IOs \\
          \hline
      (25542) Garabedian & 14.98 & 2.443 & 0.116 & 0.1224 & pro.$^{[6]}$ & IOs \\
        \hline
        \hline
    \end{tabular}
    \tablefoot{H - absolute magnitude (from JPL), proper elements are from \cite{Novakovic2017Synthetic}: $a$ – proper semi-major axis; $e$ – proper eccentricity; sin($i$) – sine of proper inclination. Notes: fugitive -- fugitive population; low-i - low inclination population; IOs - inner other population.}
    \tablebib{$^{[1]}$\cite{Durech2020ATLAS},
    $^{[2]}$\cite{Oszkiewicz2023Spins},
    $^{[3]}$\cite{Hanus2016new},
    $^{[4]}$\cite{oszkiewicz2017non},
    $^{[5]}$\cite{oszkiewicz2019physical},
    $^{[6]}$\cite{Durech2018Asteroid},
    $^{[7]}$\cite{Durech2019Inversion},
    $^{[8]}$\cite{Durech2016Asteroid}.}
\end{table}

\section{Dynamical model}
\label{dynam}
In order to apply an appropriate dynamical model to the long-term numerical integration, we adopted a very similar approach as in our previous work on the asteroid (2579) Spartacus \citep{oszkiewicz2019physical} and asteroids from the Flora region \citep{oszkiewicz2015differentiation}. The main tool was the \textsc{swift\_rmvsy} software developed by \cite{Broz2006rmvsy}, which is a modification of the \textsc{swift\_rmvs} method from the \textsc{Swift} package \citep{Duncan1998swift}.

The initial elements of the asteroids and their errors (uncertainties) with planetary data were taken from the JPL Horizons and related SBDB database\footnote{ \url{https://ssd.jpl.nasa.gov/horizons/} }. The simulation starting point was unified to JD 2459200.5 (AD 2020, December 17). The \textsc{rmvsy} method \citep{Broz2006rmvsy} was used for integration, applying the type of regularization used in the original \textsc{rmvs} algorithm \citep{Levison1994, Duncan1998swift}. This approach involves reducing the integration step when close encounters occur between massless test particles and massive bodies at distances smaller than 3 Hill radii. In practice, small bodies spend most of their integration time outside this distance range, and then the calculations are performed with a fixed-step symplectic MVS integrator, using Wisdom-Holman scheme \citep{Wisdom1991}. This procedure optimizes the computation time. In our simulations, the basic integration step was set to 10 days. A control dump of the data occurred every 1,000 years, but for presentation purposes, the data were subsequently filtered, giving the equivalent of an output of orbital elements every 100,000 years.

We used the concept of virtual clones (virtual massless test particles). To reproduce the real distribution of observational errors, we proposed the following solution: for each rotation model, we used 1001 clones of each asteroid, distributed along the variation of orbital elements. This was achieved by making the error distribution of the elements of the clones identical to that of the actual observational errors as a multidimensional Gaussian distribution. The variation of each orbital element corresponds to the original orbit determination errors (1-$\sigma$), so the dispersion of the clones in the six-dimensional space of orbital elements corresponds to the spread of the original observations. As a result, the initial conditions generated a scattered cloud in the six-element space according to the given Gaussian distribution. In the next stage, the integrator used Cartesian positions and velocities. Finally, for presentation purposes, we used the averaged proper elements that we derived from the functions and routines of the \textsc{swift\_rmvsy} package \citep{Broz2006rmvsy}. This tool contains an internal set of filtering routines, known as the 'proper-elements filter', which partially eliminates the effects of short-period perturbations related to planet-specific precession frequencies. The whole filtering procedure allowed for a better visualization of the potential impact of non-gravitational effects on the migration of selected asteroids. The cases in which clones were ejected from the system during backward integration were marginal (a few objects per thousand throughout the integration), since most of the 27 orbits studied are generally considered dynamically stable on the time scale we used. 

The cloud of clones was integrated backward for 2 billion years, with different models assuming various perturbations. This interval of time is comparable to the age of the Vesta family or slightly longer. The estimated ages of the Vesta family are related to two cratering events 1 and 2 Gy ago \citep{spoto2015}. In the simplest model (grav. model) we assumed only gravitational forces. This model includes the Sun and eight perturbing planets, with the possibility of more subtle perturbations from the largest asteroids tested in advance. As we conducted for the (2579) Spartacus object \citep{oszkiewicz2019physical} previously, here we performed a preliminary test of the extended dynamical model with the asteroids (1) Ceres, (2) Pallas, (4) Vesta, and (10) Hygiea (CPVH) on a limited time scale of $10^8$ years before the main simulation. The aim was to test whether close approaches to these asteroids could potentially affect our final results. In addition to the previously studied (2579) Spartacus,  in this way, we checked four asteroids whose elements evolve most closely to the centroid of the Vesta family: (2432) Soomana, (7899) Joya, (3536) Schleicher, and (18641) 1998 EG$_{10}$. All four asteroids have had close approaches recorded to: (1) Ceres, (2) Pallas and (4) Vesta, while only (2432) Soomana and (very occasionally) (7899) Joya approach (10) Hygiea. The cumulative effect of these approaches on the elements is detectable, but as an effect of marginal importance, we considered it negligibly small in the studied group of asteroids. The maximum differences on the main semi-major axis were on the order of $10^{-4}$ to $10^{-3}$ au on a time scale of $10^8$ years.  
Consequently, we found it reasonable to use a dynamic model consisting of the Sun and eight planets throughout the whole simulation.

In the next step, we used a more complex model with the Yarkovsky effect. Details of the thermal parameters of the asteroids used in this model \citep{Fenucci2021Role}: bulk density 3000 kg m$^{-3}$; surface density 1500 kg m$^{-3}$; surface emissivity 0.95; thermal conductivity 0.001 W K$^{-1}$ m$^{-1}$; thermal capacity of 680 J kg$^{-1}$ K$^{-1}$ for all objects and data from Table \ref{thermal_tab}.

The thermal parameters described in Table \ref{thermal_tab} are obviously only more or less accurate assumptions that we had to make in order to adopt the model for the purpose of numerical simulations. However, crucial information, such as the asteroid radius, rotation period, and direction, as well as the axis orientation, is reasonably well determined. The approximate rate and direction (positive or negative, respectively) of the Yarkovsky drift in the semi-major axis are most essential here. Therefore, our new observational results \citep{Oszkiewicz2023Spins} have the most fundamental value.

\begin{table}[h]
\caption{\label{thermal_tab}Additional thermal properties (from JPL) and rotation periods of 27 spectral V-types and V-type candidates (V$_{c}$) asteroids.}
\centering
\scriptsize\addtolength{\tabcolsep}{-2pt}
\begin{tabular}{l|c|c|c}
    \hline
    \hline
    Asteroid No. & $p_v$ & Class, & P \\
    and name & $(pV)$ & references & [h] \\
    \hline
    (956) Elisa & 0.147 $\pm$ 0.022 & V$^{[1,2,3,4,5]}$ & 16.4986 $\pm$ 0.0001 \\
    \hline
    (1914) Hartbeespoortdam & 0.212 $\pm$ 0.031  & V$^{[6,7,8,9,10]}$ & 6.339713 $\pm$ 0.00005 \\
    \hline
    (1946) Walraven & 0.362 $\pm$ 0.067 & V$^{[6,11]}$ & 10.1881 $\pm$ 0.0005 \\
    \hline
    (2432) Soomana & 0.348 $\pm$ 0.037 & V$_{c}^{[7,8,9]}$ & 3.207251 $\pm$ 0.000007 \\
    \hline
    (2566) Kirghizia & 0.264 $\pm$ 0.029 & V$^{[1,3,12,13]}$ & 4.448866 $\pm$ 0.000003 \\
    \hline
    (2579) Spartacus & 0.526 $\pm$ 0.102 & V$^{[1,3,12,13,14]}$ & 3.636030 $\pm$ 0.000002 \\
    \hline
    (2653) Principia & 0.256 $\pm$ 0.091 & V$^{[1,3,13]}$ & 5.522428 $\pm$ 0.00001 \\
    \hline
    (2704) Julian Loewe & 0.596 $\pm$ 0.065 & V$^{[1,3,7,8,9,10]}$ & 2.638333 $\pm$ 0.0000002 \\
    \hline
    (2763) Jeans & 0.412 $\pm$ 0.079 & V$^{[1,3,13,15,16,17,18]}$ & 7.801114 $\pm$ 0.0000002 \\
    \hline
    (2851) Harbin & 0.358 $\pm$ 0.026 & V$^{[1,3,12,13]}$ & 5.425895 $\pm$ 0.000002 \\
    \hline
    (2912) Lapalma & 0.456 $\pm$ 0.052 & V$^{[1,3,12,13]}$ & 5.71087 $\pm$ 0.00001 \\
    \hline
    (3307) Athabasca & 0.444 $\pm$ 0.076 & V$^{[1,3]}$ & 4.90206 $\pm$ 0.00001 \\
    \hline
    (3536) Schleicher & 0.197 $\pm$ 0.045 & V$^{[1,3,18]}$ & 5.80676 $\pm$ 0.00003 \\
    \hline
    (3849) Incidentia & 0.398 $\pm$ 0.041 & V$^{[1,3,7,8,19]}$ & 2.777072 $\pm$ 0.00002 \\
    \hline
    (4796) Lewis & 0.20~$^{*}$ & V$^{[1,3,7,8,9,13,20]}$ & 3.508471 $\pm$ 0.0000002 \\
    \hline
    (5150) Fellini & 0.419 $\pm$ 0.098 & V$_{c}^{[7,8,17,18]}$ & 5.195223 $\pm$ 0.000009 \\
    \hline
    (5524) Lecacheux & 0.034 $\pm$ 0.102 & V$_{c}^{[7,8,17,18,21]}$ & 8.41706 $\pm$ 0.00004 \\
    \hline
    (5525) 1991 TS$_{4}$ & 0.403 $\pm$ 0.046 & V$^{[7,8,9,10,22]}$ & 14.0786 $\pm$ 0.0002 \\
    \hline
    (5754) 1992 FR$_{2}$ & 0.277 $\pm$ 0.031 & V$_{c}^{[7,8,9,10,19]}$ & 8.90278 $\pm$ 0.00001 \\
    \hline
    (5952) Davemonet & 0.271 $\pm$ 0.055 & V$^{[7,8,17,18,19]}$ & 4.512535 $\pm$ 0.000004 \\
    \hline
    (6406) Mikejura & 0.512 $\pm$ 0.073 & V$^{[6,17,18,23]}$ & 6.81816 $\pm$ 0.00001 \\
    \hline
    (7558) Yurlov & 0.20~$^{*}$ & V$^{[7,8,9,10,24,25]}$ & 4.115754 $\pm$ 0.00003 \\
    \hline
    (7899) Joya & 0.555 $\pm$ 0.149 & V$_{c}^{[7,8]}$ & 85.658 $\pm$ 0.002 \\
    \hline
    (17028) 1999 FJ$_{5}$ & 0.221 $\pm$ 0.059 & V$_{c}^{[7,8]}$ & 12.80411 $\pm$ 0.00002 \\
    \hline
    (18641) 1998 EG$_{10}$ & 0.423 $\pm$ 0.147 & V$_{c}^{[7,8,9]}$ & 5.2458 $\pm$ 0.0001 \\
    \hline
    (25327) 1999 JB$_{63}$ & 0.194 $\pm$ 0.024 & V$^{[22]}$ & 6.71568 $\pm$ 0.00004 \\
    \hline
    (25542) Garabedian & 0.164 $\pm$ 0.114 & V$_{c}^{[7,8]}$ & 10.55227 $\pm$ 0.00002 \\
    \hline
    \hline
\end{tabular}
\tablefoot{$p_v$ - geometric albedo, P - rotation period. $^{*}$ data from ALCDEF \citep{Warner2019Asteroid}. }
\tablebib{$^{[1]}$\cite{Neese2002TAXONOMY}, $^{[2]}$\cite{Florczak2002Discovering}, $^{[3]}$\cite{bus2002phase}, $^{[4]}$\cite{Lazzaro2004S3OS2}, $^{[5]}$\cite{moskovitz2010spectroscopic}, $^{[6]}$\cite{alvarez2006inner}, $^{[7]}$\cite{carvano2010sdss}, $^{[8]}$\cite{SDSS2012TAXONOMY}, $^{[9]}$\cite{Veresa2015Absolute}, $^{[10]}$\cite{oszkiewicz2020spin}, $^{[11]}$\cite{ieva2016spectral}, $^{[12]}$\cite{DeMeo2009extension}, $^{[13]}$\cite{moskovitz2010spectroscopic}, $^{[14]}$\cite{oszkiewicz2019physical}, $^{[15]}$\cite{Spahr1997Discovery}, $^{[16]}$\cite{Duffard2004Mineralogical}, $^{[17]}$\cite{Licandro2017candidates}, $^{[18]}$\cite{Popescu2018Taxonomic}, $^{[19]}$\cite{hardersen2018basalt}, $^{[20]}$\cite{hasegawa2014lightcurve}, $^{[21]}$\cite{Matlovic2020Spectral}, $^{[22]}$\cite{Solontoi2012AVAST}, $^{[23]}$\cite{DeSanctis2011Spectral}, $^{[24]}$\cite{Wisniewski1991Physical}, $^{[25]}$\cite{moskovitz2008spectroscopically}.}
\end{table}

\section{Results}
\label{res}

We performed numerical integrations for a total of 27 asteroids. For each object, we investigated two symmetric rotational pole solutions from \cite{Oszkiewicz2023Spins}, then used the known diameters and geometric albedos, and assumed the remaining thermal parameters as described in Sec. \ref{dynam}. Thus, each asteroid was integrated twice with each pole solution separately. This allowed us to estimate the maximum and average values of the Yarkovsky drift that could potentially occur depending on the chosen spin model. These approximate values give a general idea of the strength of this effect and are presented in Table \ref{tab:drift}. Many of these average da/dt rates are on the order of magnitude of Yarkovsky drift that can be nowadays detected with high precision astrometry, such as that arriving from the Gaia space mission \citep{dziadura2022investigating}.

In Figs. \ref{fig:Main_1} and \ref{fig:Main_2}, we show the evolution of proper elements for all the studied asteroids within 2 Gy of backward integration. The large red points denote the current location of the investigated objects, the black points trace the evolution backward in time, and the large blue points denote their location 2 Gy ago. The current locations of the Vesta family members are marked in grey.
Figure \ref{fig:Main_1} shows the first spin orientation and Fig. \ref{fig:Main_2} shows the second spin solution. The general evolutionary trends are consistent between the two solutions for all objects.  We discuss each population separately in the following.

\subsection{Vesta fugitives}
\begin{figure}[H]
  \centering
    \includegraphics[width=1\linewidth]{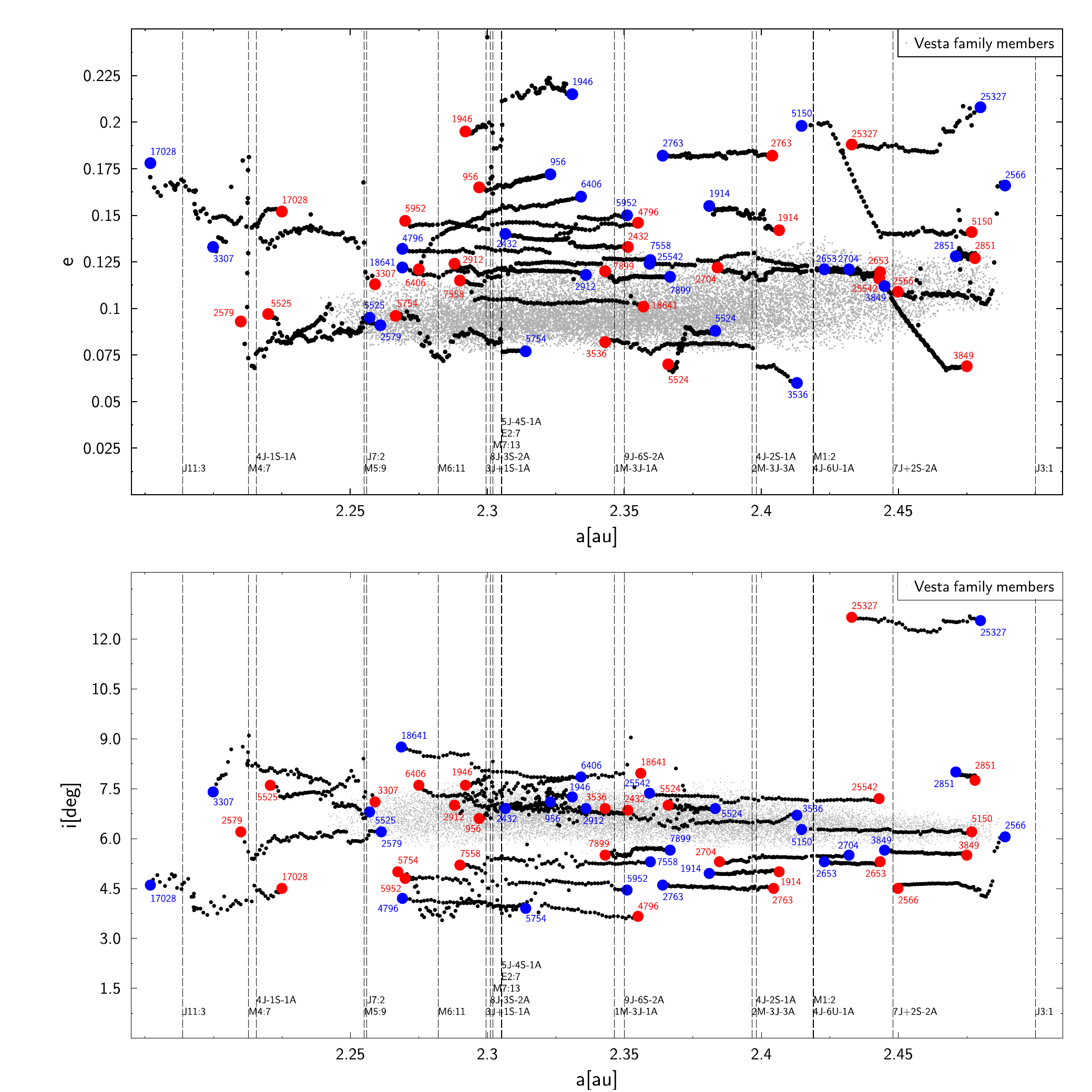}
    \caption{Evolution of proper elements of asteroids considered in this work. Black points show the long-term drift ($\Delta T$=$-$2~Gy) of proper $a$, $e$ and $i$ of the objects from Tab. \ref{tab:drift}. The red circles mark the current positions and the blue circles mark the positions of 2~Gy ago of the objects.}
  \label{fig:Main_1}
\end{figure}

\begin{figure}[H]
  \centering
    \includegraphics[width=1\linewidth]{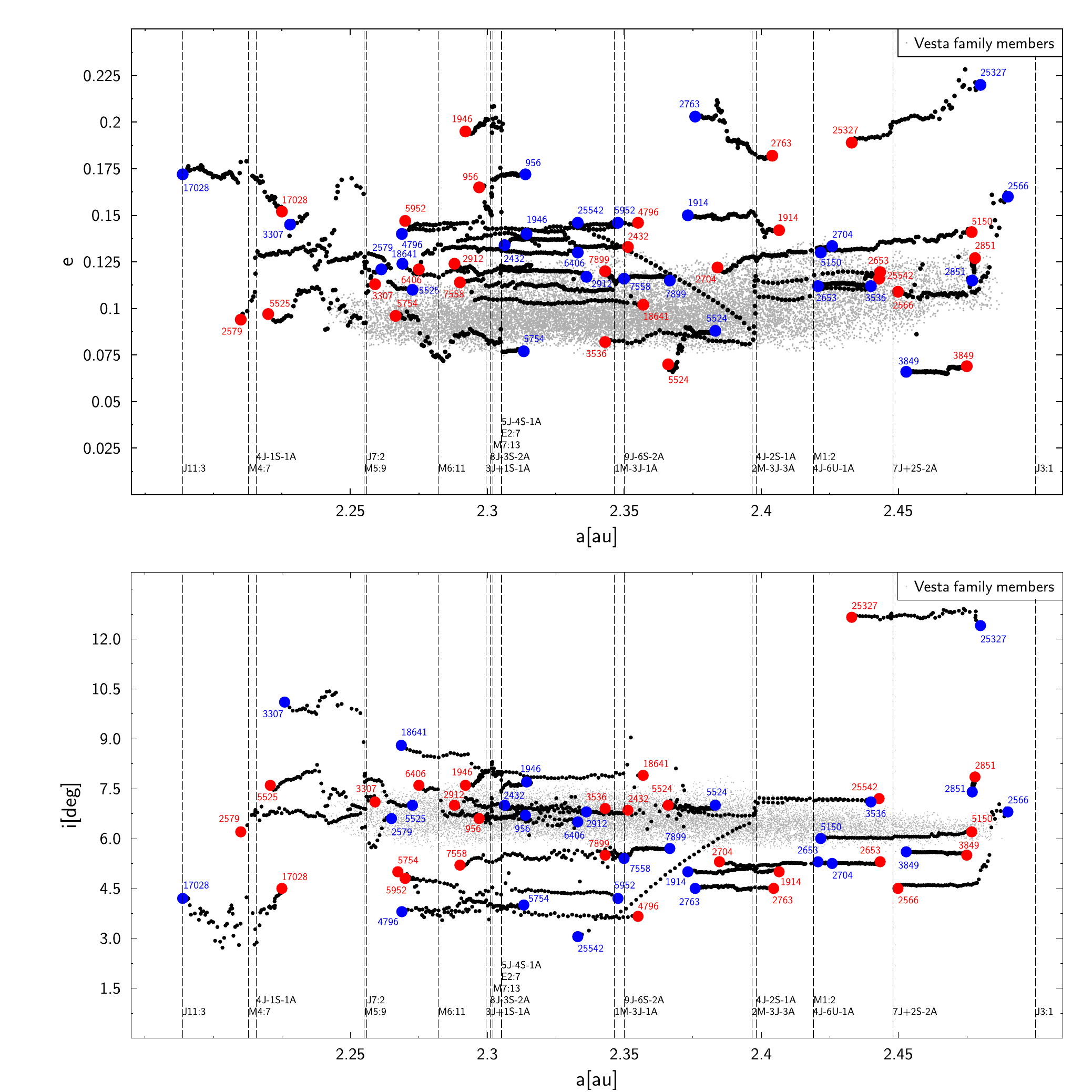}
    \caption{Evolution of proper elements of asteroids considered in this work. Same as in Fig. \ref{fig:Main_1} but for a symmetrical pole solution.}
  \label{fig:Main_2}
\end{figure}

As expected, the vast majority of our V-type asteroids can be easily linked to the core of the Vesta family within the 2 Gy time frame (e.g. (956) Elisa, (1946) Walraven, (2912) Lapalma, (5525) 1991 TS$_{4}$, (5754) 1992 FR$_{2}$, and (7558) Yurlov). Figures \ref{fig:Main_1} and \ref{fig:Main_2} show six objects that clearly overlap in the proper element phase space with the Vesta family at some point during the 2 Gy backward integration. This is consistent with previous work. \cite{nesvorny2008fugitives} performed extensive forward numerical simulations to trace the escape paths of asteroids evolving from the Vesta family. The authors found that a large fraction of objects escaped the Vesta family beyond its current boarders through the combination of dynamical mean motion resonances (e.g. the 1:2 MMR with Mars at 2.42 au, 4-2-1 MMR with Jupiter and Saturn at 2.4 au, 7:2 at 2.25 au, and three MMR at 2.3 au) and the Yarkovsky effect. Most of the evolutionary paths of the asteroids from the fugitive population investigated in this study indeed overlap in the proper orbital element phase space with that of the Vesta family members at some point during the 2 Gy integration. These objects are clearly reasonable fugitives from (4) Vesta. Our numerical simulations confirmed the results of \cite{carruba2005v} for the asteroid (956) Elisa (migrating through resonances 3-1-1, 8-3-2, and 5-4-1 MMR with Jupiter and Saturn at 2.3 au), as seen in Figs. \ref{fig:Main_1} and \ref{fig:Main_2}. As shown in our previous work, another such a fugitive is the asteroid (2579) Spartacus, which has a drift direction consistent with its origin in the the Vesta family and could have originated in an impact $\sim$ 1 Gy ago or earlier \citep{oszkiewicz2019physical}. Additional integration in this work for 2 Gy showed that the asteroid (2579) Spartacus formed in the Vesta family (see Figs. \ref{fig:Main_1} and \ref{fig:Main_2}).

Two asteroids from the fugitive population, (3307) Athabasca and (17028) 1999 FJ$_{5}$ evidently deviate from this scenario (see Figs. \ref{fig:Main_1} and \ref{fig:Main_2}). These objects are located at a small semi-major axis outside the Vesta family. Their prograde rotation translates to a negative Yarkovsky drift. The dynamical evolution of those objects clearly shows that these objects resided even further away from the Vesta family 2 Gy ago. Furthermore, extending the integration time will lead to the additional decrease of a proper semi-major axis, thus placing them even further away from the Vesta family. In our earlier work, we also found that a prograde rotator (809) Lundia is also a highly unlikely Vesta family member \citep{oszkiewicz2015differentiation}. Asteroids (3307) Athabasca and (17028) 1999 FJ$_{5}$ thus cannot be linked to (4) Vesta and may represent material left over from other differentiated planetesimals. These objects could represent the parent bodies of the Banbura Rockhole meteorite for which the origin the of the inner main belt is the most probable \citep{spurny2012bunburra, bland2009anomalous}. 

The two remaining asteroids (5952) Davemonet and (6406) Mikejura did not clearly overlap with the dynamical Vesta family during the 2~Gy integration. However, due to their location in space and passage of the orbital elements through multiple resonances around 2.3~au and 2.35~au, we do not claim that there is no link to (4) Vesta.

\subsection{Low inclination}
A number of asteroids from the low-inclination region, for instance, (1914) Hartbeespoortdam, (2653) Principia, and (2763) Jeans, (4796) Lewis) did not reach the edge of the Vesta family within the 2 Gy integration. Figures \ref{fig:Main_1} and \ref{fig:Main_2} show four such objects; the four remaining objects are traced back to the border of the family. Interestingly \cite{nesvorny2008fugitives} in their forward integration model could not reproduce the observed fraction of low-inclination V-types with sufficient efficiency within their 2 Gy integration. The authors argued that these objects could be fragments of crusts of other than (4) Vesta differentiated parent bodies. Alternatively, these bodies could have been freed from the surface of (4) Vesta during the late heavy bombardment at $\sim$ 3.9 Gy ago \citep{nesvorny2008fugitives, Scott2011Impact}.

Our work also indicates that these objects might need a longer integration time to trace their evolution back to the Vesta family. Origin in a different differentiated parent body cannot also be excluded. Intriguingly lithological differences between the vestoids and the low-$i$ asteroids were recently reported by \cite{mansour2020distribution} and differences in the median values of the 0.9$\mu$m band depth by \cite{OszkiewiczGaia2023}. Variations in mineralogy could be explained by different depths of excavation within the surface of the (4) Vesta or by the fact that the mineralogy of different bodies can be roughly the same. Additional spectral and dynamical analysis for (1914) Hartbeespoortdam, (2653) Principia, (2763) Jeans, and (4796) Lewis asteroids may help answer this question.

\subsection{Inner other}
Many of the V-types from the inner other population cannot be traced back to (4) Vesta ((2432) Soomana, (3536) Schleicher, (5150) Fellini, (5524) Lecacheux, (18641) 1998 EG$_{10}$, (25542) Garabedian). Two objects, namely, (2851) Harbin and (25327) 1999 JB$_{63}$, do not overlap with members of the Vesta family within the integration of 2 Gy in the proper-element space. The most intriguing example is asteroid (25327) 1999 JB$_{63}$, which drifts towards a larger semimajor axis and towards the 3:1 resonance when integrated backward in time.

For some of the objects studied here, the behaviour of eccentricity and inclination during long-term past evolution is similar to the changes accompanying the Kozai resonance. More precisely, there is a kind of periodic increase in eccentricity with a decrease in orbit inclination and vice versa in the case of (25542) Garabedian, which can be seen in Fig. \ref{fig:Main_2}. This asteroid has one of the largest Yarkovsky drifts (see Tab. \ref{tab:drift}). After passing the 4-2-1 MMR with Jupiter and Saturn at 2.4 au resonance, we observed periodic changes in the eccentricity and inclination of the orbit for this object. However, the time scale of these changes is inadequately large, so the presence of Kozai resonance, in this case, should be excluded.

Similar changes are accompanied by the eccentricity of the asteroid (5150) Fellini, but are not related to changes in inclination. Considering the influence of additional accelerations (drift in the semimajor axis) from the Yarkovsky effect, it is difficult to clearly separate the studied effects. In particular, when they are virtually absent for the studied asteroids in the simplest, gravitational model of forces. However, it cannot be excluded that due to nongravitational effects, they can migrate to regions where they pass through a specific kind of resonances.

Overall, among all studied asteroids, the fraction of objects that are not linked to (4) Vesta and are not explained by previous impacts is very small, indicating a small number of plausible non-Vestoids (two clear cases: (3307) Athabasca and (17028) 1999 FJ$_{5}$ in 27 studied asteroids). This is consistent with the low fraction of anomalous HEDs in the meteorite collections \citep{zhang2019oxygen}. The first of those objects was classified as V-type by \cite{bus2002phase} during the course of the Small Main-belt Asteroid Spectroscopic Survey (SMASS). Asteroid (3307) Athabasca also has a Gaia DR3 spectrum showing band depth (ratio of reflectances at 0.75$\mu$m and 0.9$\mu$m) of 1.55- a value that is more than 1$\sigma$ away from typical Vesta family members \citep{OszkiewiczGaia2023}. However, spectral follow-up observations and deeper mineralogical analysis is needed. Asteroid (17028) 1999 FJ$_{5}$ is not present in the Gaia DR3 catalogue. However, its colours, derived from the Sloan Digital Sky Survey (SDSS), are  consistent with V-type object \citep{carvano2010sdss}.

\begin{table}[h]
    \centering
\caption{\label{tab:drift}Total and mean (averaged) values of Yarkovsky drift in proper $a$, estimated numerically  with the application of two various spin axis orientations models ($\Delta T$=$-$2~Gy).}
\scriptsize\addtolength{\tabcolsep}{-2pt}
\begin{tabular}{l|c|c|c|c}
    \hline
    \hline
    Asteroid No. & $D$ & Total drift & $\Big \langle \frac{da}{dt} \Big \rangle$ & Pole coordinates \\
     and name & & $\Delta a$ & (mean drift rate) & $\lambda,\beta$ \\
     & [km] & [au] & [au/My] & [deg] \\
    \hline
    (956) Elisa & 10.474 & 2.62e-02 & 1.31e-05 & 62 $\pm$ 6, $-$54 $\pm$ 6 \\
        & $\pm$ 0.208 & 1.67e-02 & 8.35e-06 & 255 $\pm$ 4, $-$35 $\pm$ 3 \\
    \hline
    (1914) Hartbee- & 9.561 &2.29e-02 & 1.15e-05 & 37 $\pm$ 33, 29 $\pm$ 20 \\
        spoortdam & $\pm$ 0.186 & 8.29e-02 & 4.15e-05 & 175 $\pm$ 4, 52 $\pm$ 6 \\
    \hline
    (1946) Walraven & 9.205 & 7.80e-02 & 3.90e-05 & 259 $\pm$ 20, $-$80 $\pm$ 20 \\
        & $\pm$ 0.109 & 3.36e-01 & 1.68e-04 & 80 $\pm$ 20, $-$59 $\pm$ 20 \\
    \hline
    (2432) Soomana & 7.387 & 4.21e-02 & 2.10e-05 & -, 60 $\pm$ 8 \\
        & $\pm$ 0.083 & & & \\
    \hline
    (2566) Kirghizia & 7.816 & 4.30e-02 & 2.15e-05 & 98 $\pm$ 1, $-$50 $\pm$ 2 \\
        & $\pm$ 0.172 & 6.25e-02 & 3.12e-05 & 262 $\pm$ 1, $-$60 $\pm$ 4 \\
    \hline
    (2579) Spartacus & 4.604 &   6.58e-01 & 3.29e-04 & 312 $\pm$ 5, $-$57 $\pm$ 5 \\
        & $\pm$ 0.369 & 6.55e-01 & 3.28e-04 & 113 $\pm$ 5, $-$60 $\pm$ 5 \\
     \hline
    (2653) Principia & 9.882 & 1.86e-02 & 9.30e-06 & 67 $\pm$ 2, 29 $\pm$ 2 \\
        & $\pm$ 0.984 & 2.11e-02 & 1.06e-05 & 241 $\pm$ 2, 37 $\pm$ 4 \\
    \hline
    (2704) Julian Loewe & 5.199 & 4.90e-02 & 2.45e-05 & 37 $\pm$ 40, $-$88 $\pm$ 23 \\
        & $\pm$ 0.291 & 4.32e-02 & 2.16e-05 & 250 $\pm$ 50, $-$64 $\pm$ 11 \\
    \hline
    (2763) Jeans & 7.514 &   1.45e-01 & 7.24e-05   & 85 $\pm$ 5, 62 $\pm$ 5 \\
        & $\pm$ 0.157 & 1.05e+02 & 5.27e-02   & 255 $\pm$ 8, 48 $\pm$ 8 \\
    \hline
    (2851) Harbin & 8.838 & 5.72e-03 & 2.86e-06 & 99 $\pm$ 5, 10 $\pm$ 5 \\
        & $\pm$ 0.236 & 1.14e+00 & 5.68e-04 & 282 $\pm$ 5, 1 $\pm$ 5 \\
    \hline
    (2912) Lapalma & 6.519 & 4.90e-02 & 2.45e-05  & 205 $\pm$ 15, $-$75 $\pm$ 6 \\
        & $\pm$ 0.289 & & & \\
    \hline
    (3307) Athabasca & 3.628 & 5.67e-02 & 2.84e-05 & 46 $\pm$ 3, 39 $\pm$ 8 \\
        & $\pm$ 0.167 & 4.50e-01 & 2.25e-04 & 218 $\pm$ 3, 30 $\pm$ 8 \\
    \hline    
    (3536) Schleicher & 3.145 & 7.49e-02 & 3.75e-05 & 78 $\pm$ 3, $-$34 $\pm$ 6 \\
        & $\pm$ 0.161 & 1.19e-01 & 5.97e-05 & 260 $\pm$ 3, $-$51 $\pm$ 5 \\
    \hline
    (3849) Incidentia & 5.798 & 2.70e-02 & 1.35e-05 & 165 $\pm$ 6, 30 $\pm$ 9 \\
        & $\pm$ 0.125 & 1.89e-02 & 9.45e-06 & 331 $\pm$ 13, 19 $\pm$ 6 \\
    \hline
    (4796) Lewis & 5.56~$^{*}$ &  8.29e-02 & 4.15e-05  & 260 $\pm$ 52, 75 $\pm$ 15 \\
        &  &  7.26e-01 & 3.63e-04  & 328 $\pm$ 28, 81 $\pm$ 9 \\
    \hline
    (5150) Fellini & 5.401 & 5.92e-02 & 2.96e-05 & 7 $\pm$ 3, 86 $\pm$ 5 \\
        & $\pm$ 0.229 & 5.22e-02 & 2.61e-05 & 257 $\pm$ 3, 73 $\pm$ 5 \\
    \hline
    (5524) Lecacheux & 19.902 & 1.79e-02 & 8.93e-06 & -, $-$57 $\pm$ 9 \\
        & $\pm$ 12.768 & & & \\
    \hline
    (5525) 1991 TS$_{4}$ & 5.281 & 1.56e+00 & 7.80e-04 & 43 $\pm$ 50, $-$48 $\pm$ 4 \\
        & $\pm$ 0.104 & 5.43e-02 & 2.71e-05 & 223 $\pm$ 40, $-$65 $\pm$ 8 \\
    \hline
    (5754) 1992 FR$_{2}$ & 6.337 & 4.83e-02 & 2.41e-05 & 122 $\pm$ 40, $-$85 $\pm$ 11 \\
        & $\pm$ 0.078 & & & \\
    \hline
    (5952) Davemonet & 4.861 & 8.48e-02 & 4.24e-05 & 108 $\pm$ 4, $-$75 $\pm$ 4 \\
        & $\pm$ 0.275 & 8.14e-02 & 4.07e-05 & 351 $\pm$ 3, $-$73 $\pm$ 6 \\
    \hline
    (6406) Mikejura & 4.062 & 6.20e-02 & 3.10e-05 & 17 $\pm$ 5, $-$61 $\pm$ 5 \\
        & $\pm$ 0.233 & 6.09e-02 & 3.05e-05 & 216 $\pm$ 5, $-$52 $\pm$ 5 \\
    \hline
    (7558) Yurlov & 6.04~$^{*}$ &  7.24e-02  &  3.62e-05   & 121 $\pm$ 20, $-$75 $\pm$ 12 \\
        & &  6.28e-02     & 3.14e-05  & 345 $\pm$ 48, $-$64 $\pm$ 11 \\
    \hline
    (7899) Joya & 3.902 & 2.54e-02 & 1.27e-05 & -, $-$47 $\pm$ 17 \\
        & $\pm$ 0.379 & & & \\
    \hline
    (17028) 1999 FJ$_{5}$ & 1.951 & 1.85e+00 & 9.27e-04 & 40 $\pm$ 5, 18 $\pm$ 5 \\
        & $\pm$ 0.388 & 3.10e-02 & 1.55e-05 & 221 $\pm$ 5, 13 $\pm$ 5 \\
    \hline
    (18641) 1998 EG$_{10}$ & 3.717 &  8.46e-02  & 4.23e-05   & -, prograde~$^{**}$ \\
        & $\pm$ 0.650 &    &      & \\
    \hline
    (25327) 1999 JB$_{63}$ & 4.362 & 1.91e+00 & 9.54e-04 & 148 $\pm$ 2, $-$56 $\pm$ 6 \\
        & $\pm$ 0.077 & 1.58e+00 & 7.91e-04 & 339 $\pm$ 5, $-$55 $\pm$ 9 \\
    \hline
    (25542) Garabedian & 2.726 & 7.88e-02 & 3.94e-05 & 141 $\pm$ 5, 37 $\pm$ 5 \\
        & $\pm$ 0.385    & 1.05e-01 & 5.25e-05 & 312 $\pm$ 5, 61 $\pm$ 5 \\
    \hline
    \hline
\end{tabular}
\tablefoot{$^{*}$ data from ALCDEF \citep{Warner2019Asteroid}. $^{**}$ for asteroid (18641) 1998 EG$_{10}$, we only know the sense of rotation, so we use the angle at which the Yakrovsky drift is maximum.}
\end{table}

\section{Discussion}
\label{dis}

A potential additional source of uncertainty in this study is the modification of the pole orientation of the studied objects through random collisions and the Yarkovsky–O'Keefe–Radzievskii–Paddack (YORP) effect \citep{rubincam2000radiative, bottke2006yarkovsky}. The YORP effect is caused by a thermal torque and acts on small (d $<$ 40 km) asteroids that are well within the range of the objects studied in this work. The effect modifies the spin rates and spin-axis orientation of small bodies. So far, only the change in spin rates, that is, the so-called spin-up of a few asteroids, has been detected \citep{lowry2007direct, vdurech2008detection}. In the work of \cite{oszkiewicz2017non}, we have already attempted to estimate the maximum impact of the YORP effect for some asteroids, also appearing in the current study: (2704) Julian Loewe, (4796) Lewis, (5150) Fellini, (5525) 1991 TS$_{4}$, (5754) 1992 FR$_{2}$, and (18641) 1998 EG$_{10}$. In general, these considerations indicated that the axis reorientation time scales are on the same order, or longer, than the presumed lifetime of the Vesta family. Furthermore, other previously estimated obliquity rates and YORP timescales for asteroids in this work do not appear to have values large enough to significantly affect the results \citep{Golubov2021Limiting}. Therefore, we consider the Yarkovsky effect to be the most important non-gravitational force that is applicable.

It is also worth mentioning that at
this stage, the YORP effect, although present, is  hardly verifiable against the more precise and predictable Yarkovsky effect (based on a model created from our new observational data). The YORP effect contains both a static and a stochastic part, for instance, the reorientation of the axes due to collisions occurs very rarely and is supposed to have a random character. For this reason, in order to maintain better computational integrity, in the sample group, we have limited the analysis only to the simulation of the Yarkovsky effect.

It has been hypothesised before that some of the low-i V-types may have originated in collisions 3.9~Gy ago. However, such long integration times are questionable. For example, \cite{dybczynski2022important} found multiple close stellar passages near the Sun. One of those at a very close distance of around 0.014~pc ($\sim$ 3000~au) around 2.5~Mys ago. Effects of such close stellar flybys are not included in our numerical integration but could potentially introduce significant perturbations. We note that \cite{nesvorny2008fugitives} extrapolated their simulation to 3.5~Gy which also did not lead to a substantial increase of the number density of V-type fugitives in this population.

Two asteroids from the fugitive population have a prograde sense of rotation and drift in the opposite direction from the Vesta family: (3307) Athabasca and (17028) 1999 FJ$_{5}$. They show a drift to the region of the inner Solar System (2.0~au - 2.2~au). Another basaltic asteroid (908) Lundia also shows a non-Vesta drift (cannot be directly traced back to (4) Vesta) \citep{oszkiewicz2015differentiation}. 
\cite{zhang2019oxygen} based on measurements of HED meteorites (which show oxygen isotopic anomalies) \citep{bland2009anomalous, spurny2012bunburra}) suggest that there were at least five basaltic parent bodies in the past. These parent bodies may be related to objects such as (908) Lundia, (3307) Athabasca, and (17028) 1999 FJ$_{5}$, which are unlikely dynamical Vestoids. To determine the origin of these objects asteroids, further studies of their dynamical and physical properties are needed. 

A number of authors, for example \cite{Moskovitz2008distribution}, \cite{roig2008v}, \cite{Hammergren2011rotation}, \cite{Popescu2017Minor}, \cite{migliorini2017spectroscopy, Migliorini2021Characterization}, and \cite{OszkiewiczGaia2023} confirmed a number of basaltic asteroids in the middle main belt. Interestingly, asteroids (2566) Kirghizia and (25327) 1999 JB$_{63}$ show a possible drift from the 3:1 resonance at around 2.5 au. These asteroids could be V-type fragments originating from the middle main belt.

\section{Conclusions}
\label{summ}

We have investigated the dynamical evolution of 27 V-type asteroids outside the Vesta dynamical family. For the long-term dynamical integration, we used \textsc{swift\_rmvsy} software, which includes the regularized mixed-variable symplectic integration method and Yarkovsky effect. The 2~Gy backward dynamical integration took into account the physical properties of the objects, such as spin orientation, size, and thermal parameters.

Most asteroids can be explained by migration from (4) Vesta. A small fraction ($<$7\%) cannot be directly linked to (4) Vesta in our simulation. This is consistent with the low fraction of anomalous HEDs in the meteorite collections \citep{zhang2019oxygen}. Asteroids (3307) Athabasca and (17028) 1999 FJ$_{5}$ show a drift from the direction of the inner Solar System. Thus, these objects are not likely to have formed in the Vesta family. Together with a prograde rotator (908) Lundia, reported in  our earlier work \citep{oszkiewicz2015differentiation}, we have three asteroids that cannot be directly traced back to (4) Vesta. These objects could be a potential source of anomalous HED meteorites, such as the Banburra Rockhole meteorite \citep{spurny2012bunburra, bland2009anomalous}.

\section{Acknowledgments}
This work has been supported by Grant No. 2017/26/D/ST9/00240 from the National Science Center, Poland.

During the preparation of this work, the resources of the Center of Computing and Computer Modeling of the Faculty of Natural Sciences of Jan Kochanowski University in Kielce were used.

The authors thank Tomasz Kwiatkowski, Iryna Durbalova, and Antoine Choukroun for the helpful comments on our paper.

\bibliographystyle{aa}
\bibliography{biblio}

\begin{thebibliography}{86}
\expandafter\ifx\csname natexlab\endcsname\relax\def\natexlab#1{#1}\fi

\bibitem[{Alvarez-Candal {et~al.}(2006)Alvarez-Candal, Duffard, Lazzaro, \&
  Michtchenko}]{alvarez2006inner}
Alvarez-Candal, A., Duffard, R., Lazzaro, D., \& Michtchenko, T. 2006, \aap,
  459, 969

\bibitem[{Binzel \& Xu(1993)}]{binzel1993chips}
Binzel, R.~P. \& Xu, S. 1993, Sci, 260, 186

\bibitem[{Bland {et~al.}(2009)Bland, Spurn{\'y}, Towner, Bevan, Singleton,
  Bottke, Greenwood, Chesley, Shrben{\'y}, Borovi{\v{c}}ka,
  {et~al.}}]{bland2009anomalous}
Bland, P.~A., Spurn{\'y}, P., Towner, M.~C., {et~al.} 2009, Sci, 325, 1525

\bibitem[{Bottke {et~al.}(2006{\natexlab{a}})Bottke, Vokrouhlick{\`y},
  Rubincam, \& Nesvorn{\`y}}]{bottke2006yarkovsky}
Bottke, William~F., J., Vokrouhlick{\`y}, D., Rubincam, D.~P., \& Nesvorn{\`y},
  D. 2006{\natexlab{a}}, Annu. Rev. Earth Planet. Sci., 34, 157

\bibitem[{Bottke(2014)}]{bottke2014origin}
Bottke, W. 2014, in Vesta in the Light of Dawn: First Exploration of a
  Protoplanet in the Asteroid Belt, Vol. 1773, 2024

\bibitem[{Bottke {et~al.}(2006{\natexlab{b}})Bottke, Nesvorn{\'y}, Grimm,
  Morbidelli, \& O'brien}]{bottke2006iron}
Bottke, W.~F., Nesvorn{\'y}, D., Grimm, R.~E., Morbidelli, A., \& O'brien,
  D.~P. 2006{\natexlab{b}}, \nat, 439, 821

\bibitem[{Brasil {et~al.}(2017)Brasil, Roig, Nesvorn{\`y}, \&
  Carruba}]{brasil2017scattering}
Brasil, P., Roig, F., Nesvorn{\`y}, D., \& Carruba, V. 2017, \mnras, 468, 1236

\bibitem[{{Bro{\v z}}(2006)}]{Broz2006rmvsy}
{Bro{\v z}}, M. 2006, PhD thesis, Charles University in Prague

\bibitem[{Burbine {et~al.}(2001)Burbine, Buchanan, Binzel, Bus, Hiroi,
  Hinrichs, Meibom, \& McCoy}]{burbine2001vesta}
Burbine, T.~H., Buchanan, P.~C., Binzel, R.~P., {et~al.} 2001, Meteoritics \&
  Planetary Science, 36, 761

\bibitem[{Burbine {et~al.}(2002)Burbine, McCoy, Meibom, Gladman, \&
  Keil}]{burbine2002meteoritic}
Burbine, T.~H., McCoy, T.~J., Meibom, A., Gladman, B., \& Keil, K. 2002,
  Asteroids III

\bibitem[{Bus \& Binzel(2002)}]{bus2002phase}
Bus, S.~J. \& Binzel, R.~P. 2002, \icarus, 158, 106

\bibitem[{Carruba {et~al.}(2014)Carruba, Huaman, Domingos, Santos, \&
  Souami}]{carruba2014dynamical}
Carruba, V., Huaman, M., Domingos, R., Santos, C.~D., \& Souami, D. 2014,
  \mnras, 439, 3168

\bibitem[{Carruba {et~al.}(2005)Carruba, Michtchenko, Roig, Ferraz-Mello, \&
  Nesvorn{\`y}}]{carruba2005v}
Carruba, V., Michtchenko, T.~A., Roig, F., Ferraz-Mello, S., \& Nesvorn{\`y},
  D. 2005, \aap, 441, 819

\bibitem[{Carvano {et~al.}(2010)Carvano, Hasselmann, Lazzaro, \&
  Moth{\'e}-Diniz}]{carvano2010sdss}
Carvano, J., Hasselmann, P., Lazzaro, D., \& Moth{\'e}-Diniz, T. 2010, \aap,
  510, A43

\bibitem[{DeMeo {et~al.}(2009)DeMeo, Binzel, Slivan, \&
  Bus}]{DeMeo2009extension}
DeMeo, F., Binzel, R., Slivan, S., \& Bus, S. 2009, \icarus, 202, 160

\bibitem[{Duffard {et~al.}(2004)Duffard, Lazzaro, Licandro, De~Sanctis, Capria,
  \& Carvano}]{Duffard2004Mineralogical}
Duffard, R., Lazzaro, D., Licandro, J., {et~al.} 2004, \icarus, 171, 120

\bibitem[{Duffard \& Roig(2009)}]{duffard2009two}
Duffard, R. \& Roig, F. 2009, \planss, 57, 229

\bibitem[{{Duncan} {et~al.}(1998){Duncan}, {Levison}, \&
  {Lee}}]{Duncan1998swift}
{Duncan}, M.~J., {Levison}, H.~F., \& {Lee}, M.~H. 1998, AJ, 116, 2067

\bibitem[{Durech {et~al.}(2016)Durech, Hanus, Oszkiewicz, \&
  Vanco}]{Durech2016Asteroid}
Durech, J., Hanus, J., Oszkiewicz, D., \& Vanco, R. 2016, \aap, 587, A48

\bibitem[{Durech {et~al.}(2019)Durech, Hanus, \& Vanco}]{Durech2019Inversion}
Durech, J., Hanus, J., \& Vanco, R. 2019, \aap, 631, A2

\bibitem[{{\v{D}}urech {et~al.}(2008){\v{D}}urech, Vokrouhlick{\`y},
  Kaasalainen, Higgins, Krugly, Gaftonyuk, Shevchenko, Chiorny, Hamanowa,
  Reddy, {et~al.}}]{vdurech2008detection}
{\v{D}}urech, J., Vokrouhlick{\`y}, D., Kaasalainen, M., {et~al.} 2008, \aap,
  489, L25

\bibitem[{Durech {et~al.}(2020)Durech, Erasmus, Denneau, Heinze, Flewelling, \&
  Vanco}]{Durech2020ATLAS}
Durech, J. adn~Tonry, J., Erasmus, N., Denneau, L., {et~al.} 2020, \aap, 643,
  A59

\bibitem[{Dybczy{\'n}ski {et~al.}(2022)Dybczy{\'n}ski, Berski, Tokarek,
  Podlewska-Gaca, Langner, \& Bartczak}]{dybczynski2022important}
Dybczy{\'n}ski, P.~A., Berski, F., Tokarek, J., {et~al.} 2022, Astronomy \&
  Astrophysics, 664, A123

\bibitem[{Dziadura {et~al.}(2022)Dziadura, Oszkiewicz, \&
  Bartczak}]{dziadura2022investigating}
Dziadura, K., Oszkiewicz, D., \& Bartczak, P. 2022, \icarus, 383, 115040

\bibitem[{Fenucci \& Novakovic(2021)}]{Fenucci2021Role}
Fenucci, M. \& Novakovic, B. 2021, \aj, 162, id.227

\bibitem[{Florczak {et~al.}(2002)Florczak, Lazzaro, \&
  Duffard}]{Florczak2002Discovering}
Florczak, M., Lazzaro, D., \& Duffard, R. 2002, \icarus, 159, 178

\bibitem[{Golubov {et~al.}(2021)Golubov, Unukovych, \&
  Scheeres}]{Golubov2021Limiting}
Golubov, O., Unukovych, V., \& Scheeres, D.~J. 2021, \aj, 162, id.8

\bibitem[{Hammergren {et~al.}(2006)Hammergren, Gyuk, \&
  Puckett}]{hammergren200621238}
Hammergren, M., Gyuk, G., \& Puckett, A. 2006, arXiv preprint astro-ph/0609420

\bibitem[{Hammergren {et~al.}(2011)Hammergren, Gyuk, Solontoi, \&
  Puckett}]{Hammergren2011rotation}
Hammergren, M., Gyuk, G., Solontoi, M., \& Puckett, A.~W. 2011, in 42nd Lunar
  and Planetary Science Conference, Vol. 1608, 2821

\bibitem[{Hanus {et~al.}(2016)Hanus, Durech, Oszkiewicz,
  {et~al.}}]{Hanus2016new}
Hanus, J., Durech, J., Oszkiewicz, D.~A., {et~al.} 2016, \aap, 586, A108

\bibitem[{Hardersen {et~al.}(2004)Hardersen, Gaffey, \&
  Abell}]{hardersen2004mineralogy}
Hardersen, P.~S., Gaffey, M.~J., \& Abell, P.~A. 2004, \icarus, 167, 170

\bibitem[{Hardersen {et~al.}(2018)Hardersen, Reddy, Cloutis, Nowinski,
  Dievendorf, Genet, Becker, \& Roberts}]{hardersen2018basalt}
Hardersen, P.~S., Reddy, V., Cloutis, E., {et~al.} 2018, \aj, 156, 11

\bibitem[{Hasegawa {et~al.}(2014)Hasegawa, Miyasaka, Mito, Sarugaku, Ozawa,
  Kuroda, Nishihara, Harada, Yoshida, Yanagisawa,
  {et~al.}}]{hasegawa2014lightcurve}
Hasegawa, S., Miyasaka, S., Mito, H., {et~al.} 2014, Publications of the
  Astronomical Society of Japan, 66

\bibitem[{Hasselmann {et~al.}(2012)Hasselmann, Carvano, \&
  Lazzaro}]{SDSS2012TAXONOMY}
Hasselmann, P.~H., Carvano, J.~M., \& Lazzaro, D. 2012, SDSS-based Asteroid
  Taxonomy V1.1

\bibitem[{Huaman {et~al.}(2014)Huaman, Carruba, \&
  Domingos}]{huaman2014dynamical}
Huaman, M.~E., Carruba, V., \& Domingos, R.~C. 2014, \mnras, 444, 2985

\bibitem[{Ieva {et~al.}(2018)Ieva, Dotto, Lazzaro, Fulvio, Perna,
  Mazzotta~Epifani, Medeiros, \& Fulchignoni}]{ieva2018basaltic}
Ieva, S., Dotto, E., Lazzaro, D., {et~al.} 2018, \mnras, 479, 2607

\bibitem[{Ieva {et~al.}(2016)Ieva, Dotto, Lazzaro, Perna, Fulvio, \&
  Fulchignoni}]{ieva2016spectral}
Ieva, S., Dotto, E., Lazzaro, D., {et~al.} 2016, \mnras, 455, 2871

\bibitem[{Lazzaro {et~al.}(2004)Lazzaro, Angeli, Carvano, Mothe-Diniz, Duffard,
  \& Florczak}]{Lazzaro2004S3OS2}
Lazzaro, D., Angeli, C., Carvano, J., {et~al.} 2004, \icarus, 172, 179

\bibitem[{Lazzaro {et~al.}(2000)Lazzaro, Michtchenko, Carvano, Binzel, Bus,
  Burbine, Moth{\'e}-Diniz, Florczak, Angeli, \& Harris}]{lazzaro2000discovery}
Lazzaro, D., Michtchenko, T., Carvano, J., {et~al.} 2000, Sci, 288, 2033

\bibitem[{Leith {et~al.}(2017)Leith, Moskovitz, Mayne, DeMeo, Takir, Burt,
  Binzel, \& Pefkou}]{leith2017compositional}
Leith, T.~B., Moskovitz, N.~A., Mayne, R.~G., {et~al.} 2017, \icarus, 295, 61

\bibitem[{{Levison} \& {Duncan}(1994)}]{Levison1994}
{Levison}, H.~F. \& {Duncan}, M.~J. 1994, Icarus, 108, 18

\bibitem[{Licandro {et~al.}(2017)Licandro, Popescu, Morate, \&
  de~Le{\'o}n}]{Licandro2017candidates}
Licandro, J., Popescu, M., Morate, D., \& de~Le{\'o}n, J. 2017, \aap, 600, A126

\bibitem[{Lichtenberg {et~al.}(2021)Lichtenberg, Dr\c{a}{\.z}kowska,
  Sch{\"o}nb{\"a}chler, Golabek, \& Hands}]{lichtenberg2021bifurcation}
Lichtenberg, T., Dr\c{a}{\.z}kowska, J., Sch{\"o}nb{\"a}chler, M., Golabek,
  G.~J., \& Hands, T.~O. 2021, Sci, 371, 365

\bibitem[{Lowry {et~al.}(2007)Lowry, Fitzsimmons, Pravec, Vokrouhlicky,
  Boehnhardt, Taylor, Margot, Gal{\'a}d, Irwin, Irwin,
  {et~al.}}]{lowry2007direct}
Lowry, S.~C., Fitzsimmons, A., Pravec, P., {et~al.} 2007, Sci, 316, 272

\bibitem[{Mansour {et~al.}(2020)Mansour, Popescu, de~Le{\'o}n, \&
  Licandro}]{mansour2020distribution}
Mansour, J.-A., Popescu, M., de~Le{\'o}n, J., \& Licandro, J. 2020, \mnras,
  491, 5966

\bibitem[{{Marchi} {et~al.}(2012){Marchi}, {McSween}, {O'Brien}, {Schenk}, {De
  Sanctis}, {Gaskell}, {Jaumann}, {Mottola}, {Preusker}, {Raymond}, {Roatsch},
  \& {Russell}}]{Marchi2012}
{Marchi}, S., {McSween}, H.~Y., {O'Brien}, D.~P., {et~al.} 2012, Science, 336,
  690

\bibitem[{Matlovic {et~al.}(2020)Matlovic, de~Leon, Medeiros, Popescu, Rizos,
  \& Mansour}]{Matlovic2020Spectral}
Matlovic, P., de~Leon, J., Medeiros, H., {et~al.} 2020, \aap, 643, A107

\bibitem[{McCord {et~al.}(1970)McCord, Adams, \& Johnson}]{mccord1970asteroid}
McCord, T.~B., Adams, J.~B., \& Johnson, T.~V. 1970, Sci, 168, 1445

\bibitem[{McSween~Jr {et~al.}(2013)McSween~Jr, Binzel, De~Sanctis, Ammannito,
  Prettyman, Beck, Reddy, Le~Corre, Gaffey, McCord, {et~al.}}]{mcsween2013dawn}
McSween~Jr, H.~Y., Binzel, R.~P., De~Sanctis, M.~C., {et~al.} 2013, Meteoritics
  \& Planetary Science, 48, 2090

\bibitem[{Michtchenko {et~al.}(2002)Michtchenko, Lazzaro, Ferraz-Mello, \&
  Roig}]{michtchenko2002origin}
Michtchenko, T., Lazzaro, D., Ferraz-Mello, S., \& Roig, F. 2002, \icarus, 158,
  343

\bibitem[{Migliorini {et~al.}(2017)Migliorini, De~Sanctis, Lazzaro, \&
  Ammannito}]{migliorini2017spectroscopy}
Migliorini, A., De~Sanctis, M., Lazzaro, D., \& Ammannito, E. 2017, \mnras,
  475, 353

\bibitem[{Migliorini {et~al.}(2021)Migliorini, De~Sanctis, Michtchenko,
  Lazzaro, Barbieri, Mesa, \& Lazzarin}]{Migliorini2021Characterization}
Migliorini, A., De~Sanctis, M.~C., Michtchenko, T.~A., {et~al.} 2021, \mnras,
  504, 2019

\bibitem[{Moskovitz {et~al.}(2008{\natexlab{a}})Moskovitz, Jedicke, Gaidos,
  MarkWillman, Nesvorn{\'y}, RonaldFevig, \& \v{Z}eljko
  Ivezi{\'c}}]{Moskovitz2008distribution}
Moskovitz, N.~A., Jedicke, R., Gaidos, E., {et~al.} 2008{\natexlab{a}},
  \icarus, 198, 77

\bibitem[{Moskovitz {et~al.}(2008{\natexlab{b}})Moskovitz, Lawrence, Jedicke,
  Willman, Haghighipour, Bus, \& Gaidos}]{moskovitz2008spectroscopically}
Moskovitz, N.~A., Lawrence, S., Jedicke, R., {et~al.} 2008{\natexlab{b}},
  \apjl, 682, L57

\bibitem[{Moskovitz {et~al.}(2010)Moskovitz, Willman, Burbine, Binzel, \&
  Bus}]{moskovitz2010spectroscopic}
Moskovitz, N.~A., Willman, M., Burbine, T.~H., Binzel, R.~P., \& Bus, S.~J.
  2010, \icarus, 208, 773

\bibitem[{Neese(2010)}]{Neese2002TAXONOMY}
Neese, C., E. 2010, ASTEROID TAXONOMY V6.0

\bibitem[{Nesvorn{\`y}(2015)}]{Nesvorny2015HCM}
Nesvorn{\`y}, D. 2015, Nesvorny HCM Asteroid Families V3.0,
  eAR-A-VARGBDET-5-NESVORNYFAM-V3.0

\bibitem[{Nesvorn{\'y} {et~al.}(2015)Nesvorn{\'y}, Broz, \&
  Carruba}]{Nesvorny2015Asteroid}
Nesvorn{\'y}, D., Broz, M., \& Carruba, V. 2015, Asteroids IV (Univ. of Arizona
  Press, Tucson), 297--321

\bibitem[{Nesvorn{\'y} {et~al.}(2008)Nesvorn{\'y}, Roig, Gladman, Lazzaro,
  Carruba, \& Moth{\'e}-Diniz}]{nesvorny2008fugitives}
Nesvorn{\'y}, D., Roig, F., Gladman, B., {et~al.} 2008, \icarus, 193, 85

\bibitem[{Novakovic {et~al.}(2017)Novakovic, Knezevic, \&
  Milani}]{Novakovic2017Synthetic}
Novakovic, B., Knezevic, Z., \& Milani, A. 2017, Synthetic proper elements,
  524214 numbered-multiopp ast.

\bibitem[{Oszkiewicz {et~al.}(2015)Oszkiewicz, Kankiewicz, W{\l}odarczyk, \&
  Kryszczy{\'n}ska}]{oszkiewicz2015differentiation}
Oszkiewicz, D., Kankiewicz, P., W{\l}odarczyk, I., \& Kryszczy{\'n}ska, A.
  2015, \aap, 584, A18

\bibitem[{Oszkiewicz {et~al.}(2023{\natexlab{a}})Oszkiewicz, Klimczak, Carry,
  Penttil\"{a}, Popescu, Kruger, \& Keniger}]{OszkiewiczGaia2023}
Oszkiewicz, D., Klimczak, H., Carry, B., {et~al.} 2023{\natexlab{a}}, \mnras,
  519, 2917

\bibitem[{Oszkiewicz {et~al.}(2019)Oszkiewicz, Kryszczynska, Kankiewicz,
  Moskovitz, Skiff, Troianskyi, \& F{\"o}hring}]{oszkiewicz2019physical}
Oszkiewicz, D., Kryszczynska, A., Kankiewicz, P., {et~al.} 2019, \aap, 623,
  A170

\bibitem[{Oszkiewicz {et~al.}(2020)Oszkiewicz, Troianskyi, F{\"o}hring,
  Gal{\'a}d, Kwiatkowski, Marciniak, Skiff, Geier, Borczyk, Moskovitz,
  {et~al.}}]{oszkiewicz2020spin}
Oszkiewicz, D., Troianskyi, V., F{\"o}hring, D., {et~al.} 2020, \aap, 643, A117

\bibitem[{Oszkiewicz {et~al.}(2023{\natexlab{b}})Oszkiewicz, Troianskyi, Galad,
  Hanus, Durech, Wilawer, Marciniak, \& et~al.}]{Oszkiewicz2023Spins}
Oszkiewicz, D., Troianskyi, V., Galad, A., {et~al.} 2023{\natexlab{b}},
  \icarus, accepted,

\bibitem[{Oszkiewicz {et~al.}(2017)Oszkiewicz, Skiff, Moskovitz, Kankiewicz,
  Marciniak, Licandro, Galiazzo, \& Zeilinger}]{oszkiewicz2017non}
Oszkiewicz, D.~A., Skiff, B.~A., Moskovitz, N., {et~al.} 2017, \aap, 599, A107

\bibitem[{Popescu {et~al.}(2018)Popescu, Licandro, Carvano, Stoicescu,
  de~Le{\'o}n, Morate, Boac{a}, \& Cristescu}]{Popescu2018Taxonomic}
Popescu, M., Licandro, J., Carvano, J.~M., {et~al.} 2018, \aap, 617, A12

\bibitem[{Popescu {et~al.}(2017)Popescu, Licandro, Morate, de~Leon, \&
  Nedelcu}]{Popescu2017Minor}
Popescu, M., Licandro, J., Morate, D., de~Leon, J., \& Nedelcu, D.~A. 2017, The
  Messenger, 167, 16

\bibitem[{Roig {et~al.}(2008)Roig, Nesvorn{\`y}, Gil-Hutton, \&
  Lazzaro}]{roig2008v}
Roig, F., Nesvorn{\`y}, D., Gil-Hutton, R., \& Lazzaro, D. 2008, \icarus, 194,
  125

\bibitem[{Rubincam(2000)}]{rubincam2000radiative}
Rubincam, D.~P. 2000, \icarus, 148, 2

\bibitem[{Sanctis {et~al.}(2011)Sanctis, Migliorini, Jasmin, Lazzaro,
  Filacchione, Marchi, Ammannito, \& Capria}]{DeSanctis2011Spectral}
Sanctis, M. C.~D., Migliorini, A., Jasmin, F.~L., {et~al.} 2011, \aap, 533, A77

\bibitem[{Schenk {et~al.}(2012)Schenk, O’Brien, Marchi, Gaskell, Preusker,
  Roatsch, Jaumann, Buczkowski, McCord, McSween,
  {et~al.}}]{schenk2012geologically}
Schenk, P., O’Brien, D.~P., Marchi, S., {et~al.} 2012, Sci, 336, 694

\bibitem[{Schenk {et~al.}(2022)Schenk, Neesemann, Marchi, Otto, Hoogenboom,
  O’Brien, Castillo-Rogez, Raymond, \& Russell}]{schenk2022young}
Schenk, P.~M., Neesemann, A., Marchi, S., {et~al.} 2022, Meteoritics \&
  Planetary Science, 57, 22

\bibitem[{Scott \& Bottke(2011)}]{Scott2011Impact}
Scott, E. R.~D. \& Bottke, W.~F. 2011, Meteoritics and Planetary Science, 46,
  1878

\bibitem[{Scott {et~al.}(2015)Scott, Keil, Goldstein, Asphaug, Bottke, \&
  Moskovitz}]{Scott2015Asteroid}
Scott, E. R.~D., Keil, K., Goldstein, J.~I., {et~al.} 2015, Asteroids IV (Univ.
  of Arizona Press, Tucson), 573--595

\bibitem[{Solontoi {et~al.}(2012)Solontoi, Hammergren, Gyuk, \&
  AndrewPuckett}]{Solontoi2012AVAST}
Solontoi, M.~R., Hammergren, M., Gyuk, G., \& AndrewPuckett. 2012, \icarus,
  220, 577

\bibitem[{Spahr {et~al.}(1997)Spahr, Hergenrother, Larson, Hicks, Marsden,
  Williams, Tholen, Whiteley, \& Osip}]{Spahr1997Discovery}
Spahr, T.~B., Hergenrother, C.~W., Larson, S.~M., {et~al.} 1997, \icarus, 129,
  415

\bibitem[{{Spoto} {et~al.}(2015){Spoto}, {Milani}, \& {Kne{\v
  z}evi{\'c}}}]{spoto2015}
{Spoto}, F., {Milani}, A., \& {Kne{\v z}evi{\'c}}, Z. 2015, \icarus, 257, 275

\bibitem[{Spurn{\'y} {et~al.}(2012)Spurn{\'y}, Bland, Shrben{\'y},
  Borovi{\v{c}}ka, Ceplecha, Singelton, Bevan, Vaughan, Towner, McCLAFFERTY,
  {et~al.}}]{spurny2012bunburra}
Spurn{\'y}, P., Bland, P.~A., Shrben{\'y}, L., {et~al.} 2012, Meteoritics \&
  Planetary Science, 47, 163

\bibitem[{Thomas {et~al.}(1997)Thomas, Binzel, Gaffey, Storrs, Wells, \&
  Zellner}]{thomas1997impact}
Thomas, P.~C., Binzel, R.~P., Gaffey, M.~J., {et~al.} 1997, Science, 277, 1492

\bibitem[{\v{D}urech {et~al.}(2018)\v{D}urech, Hanu\v{s}, \&
  Al{\'i}-Lagoa}]{Durech2018Asteroid}
\v{D}urech, J., Hanu\v{s}, J., \& Al{\'i}-Lagoa, V. 2018, \aap, 617, A58

\bibitem[{Veres {et~al.}(2015)Veres, Jedicke, Fitzsimmons, Denneau, Granvik,
  Bolin, Chastel, Wainscoat, Burgett, Chambers, Flewelling, Kaiser, Magnier,
  Morgan, Price, Tonry, \& Waters}]{Veresa2015Absolute}
Veres, P., Jedicke, R., Fitzsimmons, A., {et~al.} 2015, \icarus, 261, 34

\bibitem[{Warner {et~al.}(2009)Warner, Harris, \& Pravec}]{Warner2019Asteroid}
Warner, B., Harris, A., \& Pravec, P. 2009, \icarus, 202, 134

\bibitem[{{Wisdom} \& {Holman}(1991)}]{Wisdom1991}
{Wisdom}, J. \& {Holman}, M. 1991, Astrophysical Journal, 102, 1528

\bibitem[{Wisniewski(1991)}]{Wisniewski1991Physical}
Wisniewski, W.~Z. 1991, \icarus, 90, 117

\bibitem[{Zhang {et~al.}(2019)Zhang, Miao, \& He}]{zhang2019oxygen}
Zhang, C., Miao, B., \& He, H. 2019, \planss

\end{thebibliography}

\onecolumn

\twocolumn

\medskip


\appendix


\end{document}